\documentclass[11pt]{article}
\usepackage{bbold}
\usepackage{geometry}
\usepackage{amsmath}

\DeclareMathOperator{\arctanh}{arctanh}

\usepackage{amsfonts}
\usepackage{amssymb}
\usepackage{graphicx}
\usepackage{caption}
\usepackage{subcaption}
\usepackage{hyperref}
\usepackage{float}
\usepackage{braket}
\usepackage{mathrsfs}
\usepackage{fancyhdr}

\title{\vspace{-1cm}\textbf{Exotic entanglement for non-Hermitian Jaynes-Cummings Hamiltonians }}
\date{}
\author{\textbf{Thomas Frith}\\\textit{Department of Mathematics, City, University of London,}\\
	\textit{Northampton Square, London EC1V 0HB, UK}\\
	\textit{E-mail: thomas.frith@city.ac.uk}}
\numberwithin{equation}{section}

\begin{document}

\maketitle 
\thispagestyle{fancy}
ABSTRACT: We provide the first solution of a time-dependent metric operator for the non-Hermitian Jaynes-Cummings Hamiltonian. We use this solution to calculate the entanglement between two identical isolated such Hamiltonians. The presence of a non-Hermitian interaction term leads to a spontaneously broken $\mathcal{PT}$-symmetric regime which manifests itself in the exotic time-evolution of entanglement. When the symmetry is broken, oscillatory modes transition into decay. As such that there is a drastic difference in behaviour between the broken and unbroken regimes.

\section{Introduction}

The Jaynes-Cummings (JC) Hamiltonian \cite{jaynes1963comparison} is a powerful model describing a two-level atom interacting with a quantized bosonic field. Being such an elegantly simple example of radiation-matter interaction, it has numerous applications is many areas of both experimental and theoretical physics. These extend to atomic and molecular physics \cite{tavis1968exact,sukumar1981multi,phoenix1991establishment,dukelsky2004exactly}, quantum optics \cite{rempe1987observation,raizen1989normal}, solid state physics \cite{hartmann2006strongly,greentree2006quantum,angelakis2007photon} and quantum information processing \cite{blais2007quantum,fink2008climbing}. There have also been numerous investigations into the non-Hermitian versions of the JC model \cite{ghosh2005exactly,bagarello2015non,bagarello2016non,brihaye2006extended}. However, these have been restricted to time-independent treatment and have therefore been unable to capture the dynamics of the system and to account for the entire parameter range. In this manuscript we present the first example of a time-dependent solution to a non-Hermitian, $\mathcal{PT}$-symmetric JC Hamiltonian with an exceptional point. Numerous studies have been conducted on $\mathcal{PT}$-symmetric non-Hermitian quantum mechanics with spontaneously broken symmetry e.g. \cite{bender1998real,bender1999pt,znojil2001spontaneous,levai2002interplay,giorgi2010spontaneous,bagchi2001generalized,dorey2001supersymmetry} and in classical $\mathcal{PT}$-symmetric optics, e.g. \cite{guo2009observation,miroshnichenko2011nonlinearly,chong2011p}. However, until recently \cite{ExactSols,FRING20172318,fring2019eternal}, the dynamics of the spontaneously broken regime was obscure. This solution includes a new time-dependent metric operator that allows us to make sense of the system when the $\mathcal{PT}$-symmetry is broken. Further to this, we make use of the metric to calculate the entanglement concurrence between two non-Hermitian JC Hamiltonians.

The JC Hamiltonian has been studied extensively in the context of entanglement e.g. \cite{phoenix1991establishment,sainz2007entanglement,bashkirov2008atom}, as it exhibits many interesting features, such as the so called "sudden death" \cite{yu2009sudden,yonacc2006sudden} of entanglement. In this manuscript we present new exotic features of entanglement between two non-Hermitian JC Hamiltonians that arise from the spontaneous breaking of $\mathcal{PT}$-symmetry within the model. 

The non-Hermitian JC Hamiltonian studied in this manuscript is
\begin{equation}\label{JCH}
H=\omega a^\dagger a+\frac{1}{2}\nu\sigma_z+ig\frac{1}{2}\left(a\sigma_++a^\dagger\sigma_-\right),
\end{equation}
with $\omega, \nu \in \Re^+$ and $g\in\Re$. $a^\dagger, a$ and $\sigma_\pm$ are the bosonic and spin creation and annihilation operators respectively. This difference between (\ref{JCH}) and the ordinary JC Hamiltonian is the existence of an imaginary interaction term $ig$. Whilst (\ref{JCH}) is non-Hermitian, it is in fact $\mathcal{PT}$-symmetric under separate transformations, either for the atom,

\begin{eqnarray}
\mathcal{PT}_{atomic}: \quad a,a^\dagger\rightarrow a,a^\dagger, \quad \sigma_\pm\rightarrow -\sigma_\pm, \quad \sigma_z\rightarrow \sigma_z, \quad i\rightarrow -i,\label{PTSymmetry1}
\end{eqnarray}
or the bosonic field,
\begin{eqnarray}
\mathcal{PT}_{bosonic}: \quad a,a^\dagger\rightarrow -a,-a^\dagger, \quad \sigma_\pm\rightarrow \sigma_\pm, \quad \sigma_z\rightarrow \sigma_z, \quad i\rightarrow -i.\label{PTSymmetry2}
\end{eqnarray}
The $\mathcal{PT}_{atomic}$ symmetry can be seen in \cite{castro2009spin}. The JCH (\ref{JCH}) has the energy eigenvalues

\begin{equation}\label{Energy}
E^\pm_n=\omega\left(n+\frac{1}{2}\right)\pm\frac{1}{2}\Omega_{n+1},
\end{equation}

with ground state
\begin{equation}
E_g=-\frac{\nu}{2},
\end{equation}

and 
\begin{equation}\label{frequencyeigen}
\Omega_{n+1}=\sqrt{\left(\omega-\nu\right)^2-\left(n+1\right)g^2},
\end{equation}
which are real for $|\left(\omega-\nu\right)|>|g\sqrt{n+1}|$. When $|\left(\omega-\nu\right)|\leq|g\sqrt{n+1}|$ the energy eigenvalues become complex and the $\mathcal{PT}$-symmetries described by (\ref{PTSymmetry1}) or (\ref{PTSymmetry2}) are spontaneously broken. This can be seen when looking at the corresponding normalised eigenstates, obtained here for the first time,

\begin{eqnarray}
\ket{\psi_n^+}&=&\frac{\cosh\left(\alpha/2\right)}{\sqrt{\cosh\left(\alpha\right)}}\ket{\uparrow,n}+i\frac{\sinh\left(\alpha/2\right)}{\sqrt{\cosh\left(\alpha\right)}}\ket{\downarrow,n+1},\\
\ket{\psi_n^-}&=&i\frac{\sinh\left(\alpha/2\right)}{\sqrt{\cosh\left(\alpha\right)}}\ket{\uparrow,n}-\frac{\cosh\left(\alpha/2\right)}{\sqrt{\cosh\left(\alpha\right)}}\ket{\downarrow,n+1},\\
\ket{\psi_g}&=&\ket{\downarrow,0},
\end{eqnarray}
with the angle $\alpha$ defined as

\begin{equation}\label{angle}
\alpha=\arctanh{\left(\frac{g\sqrt{n+1}}{\omega-\nu}\right)},
\end{equation}
and the states $\ket{\uparrow,n}$ and $\ket{\downarrow,n+1}$ denoting the combination of the photonic and spin state of the system, e.g. $\ket{\uparrow,n}=\ket{\uparrow}\otimes\ket{n}$.
The angle (\ref{angle}) becomes complex when $|\left(\omega-\nu\right)|\leq|g\sqrt{n+1}|$ leading the eigenstates to be non-normalisable. This coincides with the energy eigenvalues (\ref{Energy}) becoming complex and is the regime of spontaneously broken $\mathcal{PT}$-symmetry. It is said to be spontaneously broken as the Hamiltonian remains $\mathcal{PT}$-symmetric, but the eigenstates are broken. It is this feature that leads to the exotic entanglement described later in this manuscript.

In the field of non-Hermitian quantum mechanics, it is vital to define a metric operator, $\rho$ and a Dyson operator $\eta$ related as $\rho:=\eta^\dagger\eta$ \cite{mostafazadeh2002pseudo,bender2002complex}. The metric allows one to form a well-defined inner product $\bra{-}\rho\ket{-}$ in terms of the wavefunctions $\ket{\psi}$ of the non-Hermitian Hamiltonian. Futhermore, the Dyson operator provides a mapping to an equivalent Hermitian Hamiltonian and associated wavefunctions $\ket{\phi}$, $\ket{\phi}=\eta\ket{\psi}$. The mapping takes the form of the time-dependent Dyson equation

\begin{equation}\label{TDDE}
h\left(t\right)=\eta\left(t\right)H\eta^{-1}\left(t\right)+i\dot{\eta}\left(t\right)\eta^{-1}\left(t\right),
\end{equation}
where $h\left(t\right)$ is the associated Hermitian Hamiltonian and the overdot denotes the time-derivative. It is this equation we will solve in order to calculate $\eta\left(t\right)$. We are then able to construct $\rho\left(t\right)$ from the definition $\rho\left(t\right):=\eta^\dagger\left(t\right)\eta\left(t\right)$. We recall that operators and observables in the non-Hermitian setting are related to the Hermitian setting via a similarity transform utilising the Dyson map \cite{mostafazadeh2010pseudo}

\begin{equation}\label{Observable}
o=\eta O\eta^{-1}.
\end{equation}
This is of particular importance when we come to consider the density matrix $\varrho$. This too has an Hermitian and non-Hermitian counterpart related via a similarity transform. In the non-Hermitian setting the density matrix must include the metric operator \cite{fring2019eternal}. \\

This manuscript is organised as follows: In section \ref{TIM} we demonstrate the limitations of a time-independent metric and Dyson operator when attempting to analyse the spontaneously broken $\mathcal{PT}$ regime. This calculation will also be used to inform our ansatz for the time-dependent metric calculated in section \ref{TDM}. Finally, in section \ref{EE} we use the time-dependent metric to calculate the concurrence of entanglement for two JC Hamiltonians across a wide array of parameter sets, including the spontaneously broken $\mathcal{PT}$ regime.

\section{Time-independent mapping}\label{TIM}

In this section, we provide a new solution for Dyson operator for the non-Hermitian JC Hamiltonian. In the time-independent setting, the time-dependent Dyson equation (\ref{TDDE}) reduces to a similarity transform,

\begin{equation}
h=\eta H\eta^{-1}.
\end{equation}
The Hamiltonian (\ref{JCH}) cannot be expressed in terms of generators of a closed algebra, such an algebra would accelerate the solution procedure as it allows one to write down an ansatz for the Dyson operator in terms of generators. Instead, we begin the search for a time-independent Dyson operator by using perturbation theory \cite{scholtz1992quasi,de2006time}. In this case the Dyson operator takes the form

\begin{equation}
\eta=e^q, \quad q=\sum_{n=1}g^{2n-1}q_{2n-1},
\end{equation}
where the component $q_i$'s are unknown operators. The Hamiltonian is separated into a Hermitian and non-Hermitian component

\begin{equation}
H=H_0+iH_1,
\end{equation}
where

\begin{equation}
H_0=\omega a^\dagger a+\frac{1}{2}\nu\sigma_z, \quad H_1=g\frac{1}{2}\left(a^\dagger\sigma_-+a\sigma_+\right).
\end{equation}

We now can proceed to solve for $q_1, q_3, ...$ using the perturbation theory framework

\begin{eqnarray}
\left[H_0,q_1\right]&=&\frac{2i}{g}H_1,\\
\left[H_0,q_3\right]&=&\frac{i}{6g}\left[q_1\left[q_1,H_1\right]\right],\\
\left[H_0,q_5\right]&=&\frac{i}{6g}\left(\left[q_1\left[q_3,H_1\right]\right]+\left[q_3\left[q_1,H_1\right]\right]+\frac{1}{60}\left[q_1\left[q_1\left[q_1\left[q_1,H_1\right]\right]\right]\right]\right).
\end{eqnarray}
We solve these equations sequentially and obtain the following expressions

\begin{eqnarray}
q_1&=&\frac{i}{\left(\omega-\nu\right)}\left(a^\dagger\sigma_--a\sigma_+\right),\\
q_3&=&\frac{i}{3\left(\omega-\nu\right)^3}\left(a^\dagger a a^\dagger\sigma_--aa^\dagger a\sigma_+\right),\\
q_5&=&\frac{i}{5\left(\omega-\nu\right)^5}\left(a^\dagger a a^\dagger a a^\dagger\sigma_--aa^\dagger a a^\dagger a\sigma_+\right).
\end{eqnarray}
The series continues in a fashion that we can predict, in fact we can write down a closed expression for $q$

\begin{equation}
q=ia^\dagger\left(aa^\dagger\right)^{-1/2}\arctan\left[\frac{g \left(aa^\dagger\right)^{1/2}}{\omega-\nu}\right]\sigma_--ia\left(a^\dagger a\right)^{-1/2}\arctan\left[\frac{g\left(a^\dagger a\right)^{1/2}}{\omega-\nu}\right]\sigma_+.
\end{equation}
Perturbation theory guided us to the exact solution and now we can find the counterpart Hermitian Hamiltonian

\begin{equation}
h=\omega\left(a^\dagger a+\frac{\sigma_z}{2}\right)-\frac{1}{4}\left(I+\sigma_z\right)\Omega_{aa^\dagger}+\frac{1}{4}\left(I-\sigma_z\right)\Omega_{a^\dagger a},
\end{equation}
where 
\begin{equation}\label{frequency}
\Omega_{a^\dagger a}=\sqrt{\left(\omega-\nu\right)^2-g^2a^\dagger a}, \quad \Omega_{aa^\dagger }=\sqrt{\left(\omega-\nu\right)^2-g^2aa^\dagger },
\end{equation}
are operator versions of the frequency described in equation (\ref{frequencyeigen}). In fact, (\ref{frequencyeigen}) is the eigenvalue of $\Omega_{aa^\dagger }$. We can see this by acting on a state $\ket{n}$,
\begin{equation}\label{frequencyeq}
\Omega_{a^\dagger a}\ket{n}=\Omega_{n}\ket{n}, \quad \Omega_{aa^\dagger }\ket{n}=\Omega_{n+1}\ket{n}.
\end{equation}
We can then see that this Hermitian Hamiltonian shares the same energy eigenvalues as (\ref{JCH}). Therefore the mapping must break down when $|\omega-\nu|<|g\sqrt{n+1}|$. This is now the motivation for finding a time-dependent Dyson operator. In the time-dependent regime the form of the time-dependent Dyson equation allows for a Dyson operator and metric to exist even when the $\mathcal{PT}$-symmetry is broken and the energy eigenvalues of (\ref{JCH}) are complex.

\section{Time-dependent mapping}\label{TDM}

We now introduce a time-dependence into the Dyson operator and therefore solve equation (\ref{TDDE}) in full. This is the first time-dependent solution for the Dyson operator for the non-Hermitian JC Hamiltonian. The results from the time-independent mapping inform us of the form the Dyson map should take and so we make the ansatz
\begin{equation}
\eta\left(t\right)=e^{q_z\left(t\right)}e^{q_-\left(t\right)},
\end{equation}
where the matrix operator valued functions $q_z$ and $q_-$ are of the form

\begin{equation}
q_z\left(t\right)=\frac{1}{2}K_{aa^\dagger}\left(t\right)\left(I+\sigma_z\right)-\frac{1}{2}K_{a^\dagger a}\left(t\right)\left(I-\sigma_z\right), \quad q_-=a^\dagger\left(aa^\dagger\right)^{1/2}f_{aa^\dagger}\left(t\right)\sigma_-.
\end{equation}
$K_{aa^\dagger,a^\dagger a}$ and $f_{aa^\dagger,a^\dagger a}$ are unknown functions of time and the operators $aa^\dagger, a^\dagger a$. $K$ is real and $f$ may be complex. Acting on a state $\ket{n}$, as we did in equation (\ref{frequencyeq}), we see that the operators have eigenvalues,
\begin{equation}
K_{aa^\dagger,a^\dagger a}\ket{n}=K_{n+1,n}\ket{n}, \quad f_{aa^\dagger,a^\dagger a}\ket{n}=f_{n+1,n}\ket{n}.
\end{equation}
The adjoint actions of these operators on the components of the Hamiltonian are

\begin{eqnarray}
e^{q_-}He^{-q_-}&=&H_0+iH_1-ig\frac{1}{2}\left(aa^\dagger\right)^{1/2}f_{aa^\dagger}\left(I+\sigma_z\right)+ig\frac{1}{2}\left(a^\dagger a\right)^{1/2}f_{aa^\dagger}\left(I-\sigma_z\right)\\
&&-\left[\left(\omega-\nu\right)a^\dagger\left(a^\dagger a\right)^{1/2}f_{aa^\dagger}+iga^\dagger f_{aa^\dagger}^2\right]\sigma_-\nonumber\\ e^{q_z}e^{q_-}He^{-q_-}e^{-q_z}&=&H_0-ig\frac{1}{2}\left(aa^\dagger\right)^{1/2}f_{aa^\dagger}\left(I+\sigma_z\right)+ig\frac{1}{2}\left(a^\dagger a\right)^{1/2}f_{aa^\dagger}\left(I-\sigma_z\right)\label{adjoint}\\
&&-\left[iga^\dagger\left(\omega-\nu\right)a^\dagger\left(a^\dagger a\right)^{1/2}f_{aa^\dagger}+iga^\dagger f_{aa^\dagger}^2\right]e^{-2K_{aa^\dagger}}\sigma_-+igae^{2K_{aa^\dagger}}\sigma_+\nonumber\\
i\dot{\eta}\eta^{-1}&=&-ig\frac{1}{2}\dot{K}_{aa^\dagger}\left(I+\sigma_z\right)+ig\frac{1}{2}\dot{K}_{aa^\dagger}\left(I-\sigma_z\right)\label{timecomponent}\\
&&+ia^\dagger\left(aa^\dagger\right)^{1/2}\dot{f}_{aa^\dagger}e^{-2K_{aa^\dagger}}\nonumber.
\end{eqnarray}
With expressions (\ref{adjoint}) and (\ref{timecomponent}) we have the necessary components to evaluate the time-dependent Dyson equation (\ref{TDDE}). Finally, we write $f=\alpha+i\beta$ where $\alpha, \beta \in \Re$ and proceed to eliminate the non-Hermitian terms from 
\begin{equation}
h=\eta H \eta^{-1}+i\dot{\eta}\eta^{-1}=e^{q_z}e^{q_-}He^{-q_-}e^{-q_z}+i\dot{\eta}\eta^{-1}.
\end{equation}
Eliminating these components leads to the following constraining differential equations
\begin{eqnarray}
\dot{K}_{aa^\dagger}&=&g\frac{1}{2}\left(aa^\dagger\right)^{1/2}\alpha_{aa^\dagger},\label{Keq}\\
\dot{\alpha}_{aa^\dagger}&=&\left(\omega-\nu\right)\beta_{aa^\dagger}-g\frac{1}{2}\left(1-\alpha_{aa^\dagger}^2+\beta_{aa^\dagger}^2\right)\left(aa^\dagger\right)^{1/2}-g\frac{1}{2}\left(aa^\dagger\right)^{1/2}e^{4K_{aa^\dagger}},\label{alphaeq}\\
\dot{\beta}_{aa^\dagger}&=&-\left(\omega-\nu\right)\alpha_{aa^\dagger}+g\left(aa^\dagger\right)^{1/2}\alpha_{aa^\dagger}\beta_{aa^\dagger}\label{betaeq}.
\end{eqnarray}
These coupled differential equations can be solved systematically. Equation (\ref{Keq}) can be solved for $\alpha$
\begin{equation}
\alpha_{aa^\dagger}=\frac{1}{2}\dot{\delta}_{aa^\dagger}\left[g\left(aa^\dagger\right)^{1/2}\delta_{aa^\dagger}\right]^{-1}, \quad \delta_{aa^\dagger}=e^{2K_{aa^\dagger}}.
\end{equation}
Substituting this into equation (\ref{betaeq}) and solving for $\beta$ gives the expression
\begin{equation}
\beta_{aa^\dagger}=\frac{\omega-\nu}{g}\left(aa^\dagger\right)^{-1/2}+c_{1,{aa^\dagger}}\delta_{aa^\dagger},
\end{equation}
where we obtain the time-independent operator $c_{1,{aa^\dagger}}$. As we will see, this is defined when considering the initial conditions. Finally, substituting both $\alpha$ and $\beta$ into equation (\ref{alphaeq}) results in the second order differential equation in terms of $\delta$
\begin{equation}
\frac{\ddot{\delta}_{aa^\dagger}}{2}\delta_{aa^\dagger}^{-1}-\frac{3\dot{\delta}_{aa^\dagger}^2}{4}\delta_{aa^\dagger}^{-2}+g^2\left(1+c_{1,{aa^\dagger}}^2\right)\left(aa^\dagger\right)\delta_{aa^\dagger}^2-\frac{1}{4}\left[\left(\omega-\nu\right)^2-g^2\left(aa^\dagger\right)\right]=0.
\end{equation}
Using the substitution $\delta_{aa^\dagger}=\sigma_{aa^\dagger}^{-2}$, this reduces to the operator valued Ermokov-Pinney equation \cite{ermakov1880transformation,pinney1950nonlinear}
\begin{equation}
\ddot{\sigma}_{aa^\dagger}+\frac{1}{4}\Omega_{aa^\dagger}^2\sigma_{aa^\dagger}=g^2\left(1+c_{1,{aa^\dagger}}^2\right)\left(aa^\dagger\right)\sigma_{aa^\dagger}^{-3}.
\end{equation}
The solution to this is

\begin{equation}
\sigma_{aa^\dagger}\left(t\right)=\sqrt{c_{2,{aa^\dagger}}\cos\left(\Omega_{aa^\dagger} t+c_{3,{aa^\dagger}}\right)+c_{4,{aa^\dagger}}},
\end{equation}
with
\begin{equation}
c_{4,{aa^\dagger}}=\Omega_{aa^\dagger}^{-1}\sqrt{4\left(1+c_{1,{aa^\dagger}}^2-c_{2,{aa^\dagger}}^2\right)g^2\left(aa^\dagger\right)+c_{2,{aa^\dagger}}^2\left(\omega-\nu\right)^2}.
\end{equation}
The resulting Hermitian Hamiltonian is time-dependent and takes the form

\begin{eqnarray}
h\left(t\right)&=&H_0+g\frac{1}{2}\left(aa^\dagger\right)^{1/2}\beta_{aa^\dagger }\left(I+\sigma_z\right)-g\frac{1}{2}\left(a^\dagger a\right)^{1/2}\beta_{a^\dagger a}\left(I-\sigma_z\right)\\
&+&ig\left(a\delta_{a^\dagger a}\sigma_+-a^\dagger\delta_{aa^\dagger}\sigma_-\right).
\end{eqnarray}
In order to set the time-independent operators $c_{1,{aa^\dagger}}$, $c_{2,{aa^\dagger}}$ and $c_{3,{aa^\dagger}}$ we must fix the initial condition for $\eta\left(t\right)$. For simplicity, we choose $\eta\left(0\right)=\mathcal{I}$. This sets the operators to 

\begin{eqnarray}
c_{1,{aa^\dagger}}&=&-\frac{\omega-\nu}{g}\left(aa^\dagger\right)^{-1},\\
c_{2,{aa^\dagger}}&=&-\Omega_{aa^\dagger}^{-2}g^2\left(aa^\dagger\right),\\
c_{3,{aa^\dagger}}&=&0,
\end{eqnarray}
which sets $c_{4,{aa^\dagger}}$ to be
\begin{eqnarray}
c_{4,{aa^\dagger}}&=&\Omega_{aa^\dagger}^{-2}\left(\omega-\nu\right)^2.
\end{eqnarray}
Therefore the time-dependent, operator valued functions of the Dyson map are

\begin{eqnarray}
\delta_{aa^\dagger}\left(t\right)&=&\frac{1}{2}\ln K_{aa^\dagger}\left(t\right)=\Omega_{aa^\dagger}^2\left[\left(\omega-\nu\right)^2-g^2\left(aa^\dagger\right)\cos\left(\Omega_{aa^\dagger}t\right)\right]^{-1},\label{delta}\\
\alpha_{aa^\dagger}\left(t\right)&=&-g\left(aa^\dagger\right)^{1/2}\Omega_{aa^\dagger}\sin\left(\Omega_{aa^\dagger}t\right)\left[\left(\omega-\nu\right)^2-g^2\left(aa^\dagger\right)\cos\left(\Omega_{aa^\dagger}t\right)\right]^{-1},\\
\beta_{aa^\dagger}\left(t\right)&=&g\left(aa^\dagger\right)^{1/2}\left(\omega-\nu\right)\left(1-\cos\left(\Omega_{aa^\dagger}t\right)\right)\left[\left(\omega-\nu\right)^2-g^2\left(aa^\dagger\right)\cos\left(\Omega_{aa^\dagger}t\right)\right]^{-1}.
\end{eqnarray}

\section{Exotic entanglement}\label{EE}

We now study in detail a system consisting of two JC Hamiltonians. These two Hamiltonains are totally separated and have no communication with one another. Therefore we can write the Hamiltonian as a simple sum of two JC Hamiltonians.

\begin{equation}\label{twoJC}
H=\omega a^\dagger a+\omega b^\dagger b+\nu\sigma_z^a+\nu\sigma_z^b+ig\frac{1}{2}\left(a\sigma_+^a+a^\dagger\sigma_-^a\right)+ig\frac{1}{2}\left(b\sigma_+^b+b^\dagger\sigma_-^b\right),
\end{equation}
where we now have two sets of creation, annihilation and spin operators, introducing $b$, $b^\dagger$ and $\sigma_{\pm,z}^b$ for the second Hamiltonian. It follows that because the two Hamiltonians are totally isolated we can form the Dyson operator and metric as a simple product of the two systems $\rho=\rho_a\rho_b$, $\eta=\eta_a\eta_b$. In order to construct a wavefunction for the system (\ref{twoJC}) we must define an initial condition. We clearly wish to observe entanglement and so we ensure the initial state is entangled to some degree. Therefore we set the initial state of the atoms to be

\begin{equation}
\ket{\Psi_{Atom}}=\cos\gamma\ket{\uparrow\uparrow}+\sin\gamma\ket{\downarrow\downarrow},
\end{equation}
where the states $\ket{\uparrow\uparrow}=\ket{\uparrow_a}\otimes\ket{\uparrow_b}$ and $\ket{\downarrow\downarrow}=\ket{\downarrow_a}\otimes\ket{\downarrow_b}$.
The initial state of the cavities we choose to be $\ket{0n}=\ket{0_a}\otimes\ket{n_b}$, this will allow us to observe the wide array of exotic entanglement effects unique to non-Hermitian systems. The initial state of the whole system is therefore $\ket{\Psi_{Atom}}\otimes\ket{0n}$, 

\begin{equation}\label{initial_condition}
\ket{\psi_0}=\cos\gamma\ket{\uparrow\uparrow 0n}+\sin\gamma\ket{\downarrow\downarrow 0n}.
\end{equation}
Solving the time-dependent Schr\"odinger equation with this initial state leads to the time-dependent wavefunction

\begin{eqnarray}\label{state}
\ket{\psi\left(t\right)}&=&x_1\left(t\right)\ket{\downarrow\downarrow 0\;n}+x_2\left(t\right)\ket{\downarrow\uparrow 0\;n-1}+x_3\left(t\right)\ket{\uparrow\uparrow 0\;n}\\
&+&x_4\left(t\right)\ket{\uparrow\downarrow 0\;n+1}+x_5\left(t\right)\ket{\downarrow\uparrow 1\;n}+x_6\left(t\right)\ket{\downarrow\downarrow 1\;n+1},\nonumber
\end{eqnarray}
which is a combination of the eigenfunctions $\ket{\psi_{g,a}}$,  $\ket{\psi^{\pm}_{1,a}}$, $\ket{\psi^{\pm}_{n-1,b}}$ and $\ket{\psi^{\pm}_{n,b}}$ needed to allow the initial condition (\ref{initial_condition}). The time-dependent coefficient functions are

\begin{eqnarray}\label{NHParams}
x_1\left(t\right)&=& U^*_{n}\left(t\right)e^{-\frac{1}{2}i\left(\omega-\nu\right)t}\sin\gamma,\\
x_2\left(t\right)&=& D_n\left(t\right)e^{-\frac{1}{2}i\left(\omega-\nu\right)t}\sin\gamma,\\
x_3\left(t\right)&=& U_1\left(t\right)U_{n+1}\left(t\right)e^{-i\omega t}\cos\gamma,\\
x_4\left(t\right)&=& U_1\left(t\right)D_{n+1}\left(t\right)e^{-i\omega t}\cos\gamma,\\
x_5\left(t\right)&=& D_1\left(t\right)U_{n+1}\left(t\right)e^{-i\omega t}\cos\gamma,\\
x_6\left(t\right)&=& D_1\left(t\right)D_{n+1}\left(t\right)e^{-i\omega t}\cos\gamma.
\end{eqnarray}
The functions $U_n\left(t\right)$ and $D_n\left(t\right)$ are defined as

\begin{eqnarray}
U_n\left(t\right)&=&\left[\cos\left(\frac{1}{2}\Omega_nt\right)+\frac{i\left(\omega-\nu\right)}{\Omega_n}\sin\left(\frac{1}{2}\Omega_nt\right)\right]e^{-i\left(n-1\right)\omega t},\\
D_n\left(t\right)&=&\frac{g\sqrt{n}}{\Omega_n}\sin\left(\frac{1}{2}\Omega_nt\right)e^{-i\left(n-1\right)\omega t},
\end{eqnarray}
where we recall $\Omega_n=\sqrt{\left(\omega-\nu\right)^2-ng^2}$. Therefore we see that for $n>1$, a general state contains three frequencies, $\Omega_1$, $\Omega_n$ and $\Omega_{n+1}$. This should follow from the fact we have a general wavefunction containing the eigenfunctions $\ket{\psi_{g,a}}$,  $\ket{\psi^{\pm}_{1,a}}$, $\ket{\psi^{\pm}_{n-1,b}}$ and $\ket{\psi^{\pm}_{n,b}}$. For $n=0$ there is only one frequency at $\Omega_1$. For $n=1$ there are two frequencies at $\Omega_1$ and $\Omega_2$. \\

To study the entanglement of this system, we must construct the density matrix \cite{fring2019eternal} 

\begin{equation}\label{NHmetric}
\varrho_H\left(t\right)=\ket{\psi\left(t\right)}\bra{\psi\left(t\right)}\rho\left(t\right),
\end{equation}
and then take the similarity transform in order to calculate the Hermitian density matrix in accordance with equation (\ref{Observable})

\begin{equation}\label{Hmetric}
\varrho_h\left(t\right)=\eta\left(t\right)\varrho_H\left(t\right)\eta^{-1}\left(t\right).
\end{equation}
In order to find the density matrix we can proceed in two ways: The first is to use the wavefunction (\ref{state}) along with the metric to construct the non-Hermitian density matrix (\ref{NHmetric}). We can then use the Dyson operator to perform the similarity transform (\ref{Hmetric}) in order to calculate the Hermitian density matrix. The second approach is to use the Dyson operator to transform the wavefunction (\ref{state}) into the equivalent wavefunction for the Hermitian system $\ket{\phi\left(t\right)}=\eta\left(t\right)\ket{\psi\left(t\right)}$ (as described in the introduction). We can then calculate the Hermitian density matrix directly from this new wavefunction

\begin{equation}\label{Hmetric2}
\varrho_h\left(t\right)=\ket{\phi\left(t\right)}\bra{\phi\left(t\right)}.
\end{equation}
Both approaches are equivalent but the second approach requires far fewer calculations and therefore we proceed in this manner, calculating $\varrho_h\left(t\right)$ using equation (\ref{Hmetric2}) and $\ket{\phi\left(t\right)}$ (as was done in \cite{fring2019eternal}). Acting with $\eta\left(t\right)$ calculated in section \ref{TDM} on $\ket{\psi\left(t\right)}$ from (\ref{state}), we find the transformed wavefunction $\ket{\phi\left(t\right)}$ to be

\begin{eqnarray}\label{statetransformed}
\ket{\phi\left(t\right)}&=&y_1\left(t\right)\ket{\downarrow\downarrow 0\;n}+y_2\left(t\right)\ket{\downarrow\uparrow 0\;n-1}+y_3\left(t\right)\ket{\uparrow\uparrow 0\;n}\\
&+&y_4\left(t\right)\ket{\uparrow\downarrow 0\;n+1}+y_5\left(t\right)\ket{\downarrow\uparrow 1\;n}+y_6\left(t\right)\ket{\downarrow\downarrow 1\;n+1}.\nonumber
\end{eqnarray}
where the new time-dependent coefficient functions are constructed from those in (\ref{NHParams}) and the coefficient functions in the Dyson operator

\begin{eqnarray}
y_1\left(t\right)&=& x_1\left(t\right)\delta_n^{1/2},\\
y_2\left(t\right)&=& x_2\left(t\right)\delta_n^{1/2},\\
y_3\left(t\right)&=& x_3\left(t\right)\delta_1^{1/2}\delta_{n+1}^{1/2},\\
y_4\left(t\right)&=& -x_4\left(t\right)\delta_1^{1/2}\delta_{n+1}^{1/2},\\
y_5\left(t\right)&=& -x_5\left(t\right)\delta_1^{1/2}\delta_{n+1}^{1/2},\\
y_6\left(t\right)&=& x_6\left(t\right)\delta_1^{1/2}\delta_{n+1}^{1/2}.\\
\end{eqnarray}
We recall $\delta_{aa^\dagger}$ is defined in equation (\ref{delta}), $\delta_{n+1}$ and $\delta_{n}$ are the resulting eigenfunctions when acting on a state $\ket{n}$ with $\delta_{aa^\dagger}$ and $\delta_{a^\dagger a}$
\begin{equation}
\delta_{aa^\dagger}\ket{n}=\delta_{n+1}\ket{n}, \quad \delta_{a^\dagger a}\ket{n}=\delta_n\ket{n}.
\end{equation}

With the construction of this state, we can now form the Hermitian reduced density matrix of the two atoms by tracing out the photonic part of the wavefunction 

\begin{equation}
\varrho_h=\begin{pmatrix}
|y_3|^2 & 0 & 0 & y_3y_1^* \\
0 & |y_2|^2+|y_5|^2 & 0 & 0 \\
0 & 0 & |y_4|^2 & 0 \\
y_3^*y_1 & 0 & 0 & |y_1|^2+|y_6|^2 
\end{pmatrix}.
\end{equation}
From this density matrix we calculate the concurrence 

\begin{equation}
C\left(t\right)=\max\{0,f\left(t\right)\},
\end{equation}
using Wooters definition \cite{wootters1998entanglement}, where

\begin{equation}
f\left(t\right)=2|y_3|\sqrt{|y_1|^2+|y_6|^2}-2|y_4|\sqrt{|y_2|^2+|y_5|^2}.
\end{equation}
We now plot the concurrence of the state (\ref{state}) for $n=0$, $n=1$ and $n=2$ and note the rich array of exotic effects we find in each case. We recall that for $n=0$, the only frequency present is $\Omega_1$; for $n=1$, the frequencies are $\Omega_1$ and $\Omega_2$; for $n=2$ the frequencies are $\Omega_1$, $\Omega_2$ and $\Omega_3$. We vary the parameter $\kappa=\frac{\omega-\nu}{g}$, such that $\Omega_m=g\sqrt{\kappa^2-m}$, where $m$ is the index denoting the frequency present. This will determine whether the frequency modes are in the $\mathcal{PT}$-symmetric or broken regime; for $m<\kappa^2$, $\Omega_m$ is real and the symmetry is intact. We then plot the concurrence with respect to $g t/\pi$.

%
%
%
%
%

\begin{figure}[H]
	\centering
	\begin{subfigure}[t]{0.45\textwidth}
		\includegraphics[scale=0.55]{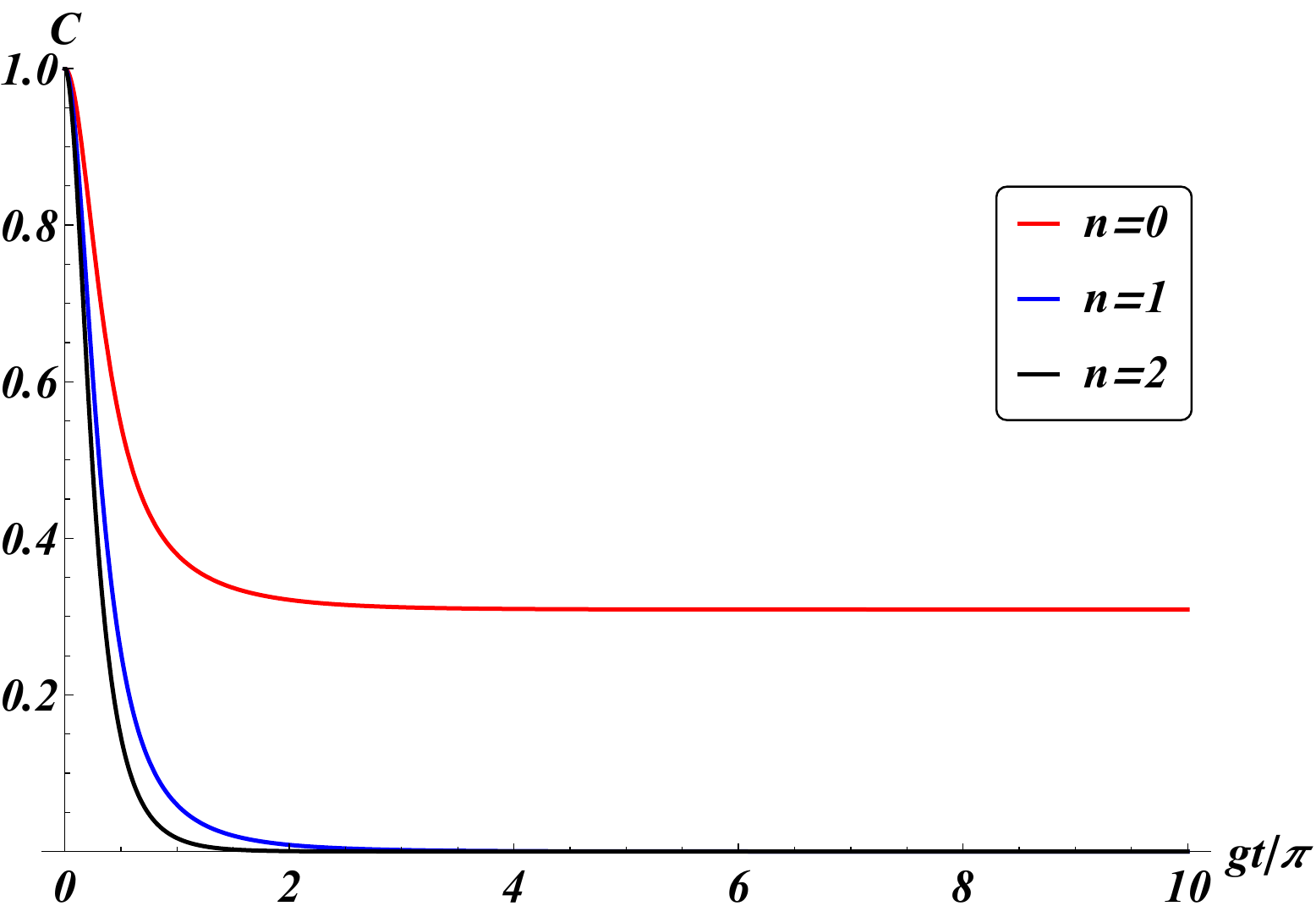}
		\caption{\scriptsize$\kappa=0.9$, all frequencies are in the broken regime.}\label{a}
	\end{subfigure}\hfill
	\begin{subfigure}[t]{0.45\textwidth}
		\includegraphics[scale=0.38]{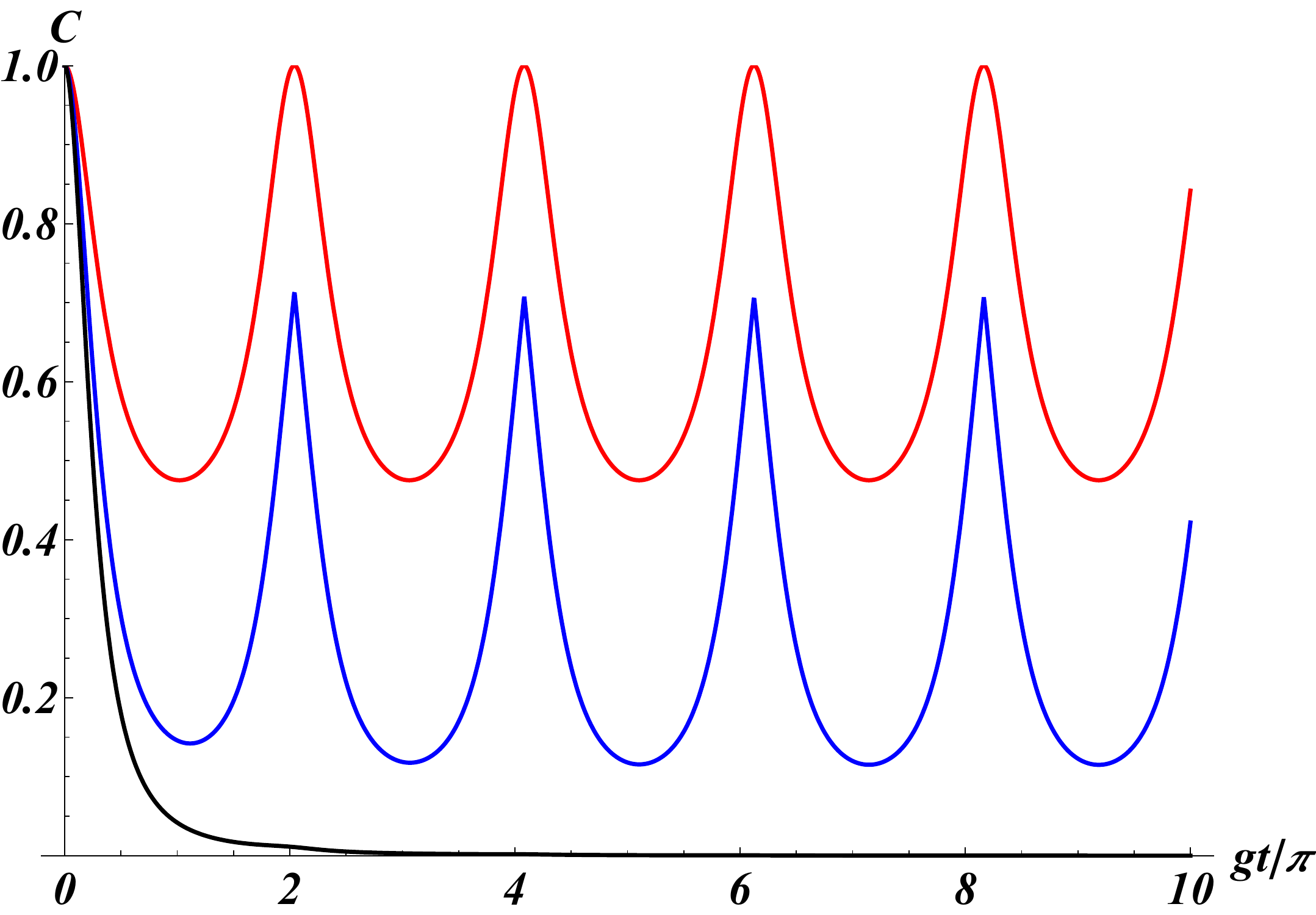}
		\caption{\scriptsize$\kappa=1.4$, $\Omega_3$ and $\Omega_2$ are in the broken regime.}\label{b}
	\end{subfigure}\vfill
	\begin{subfigure}[b]{0.45\textwidth}
		\includegraphics[scale=0.38]{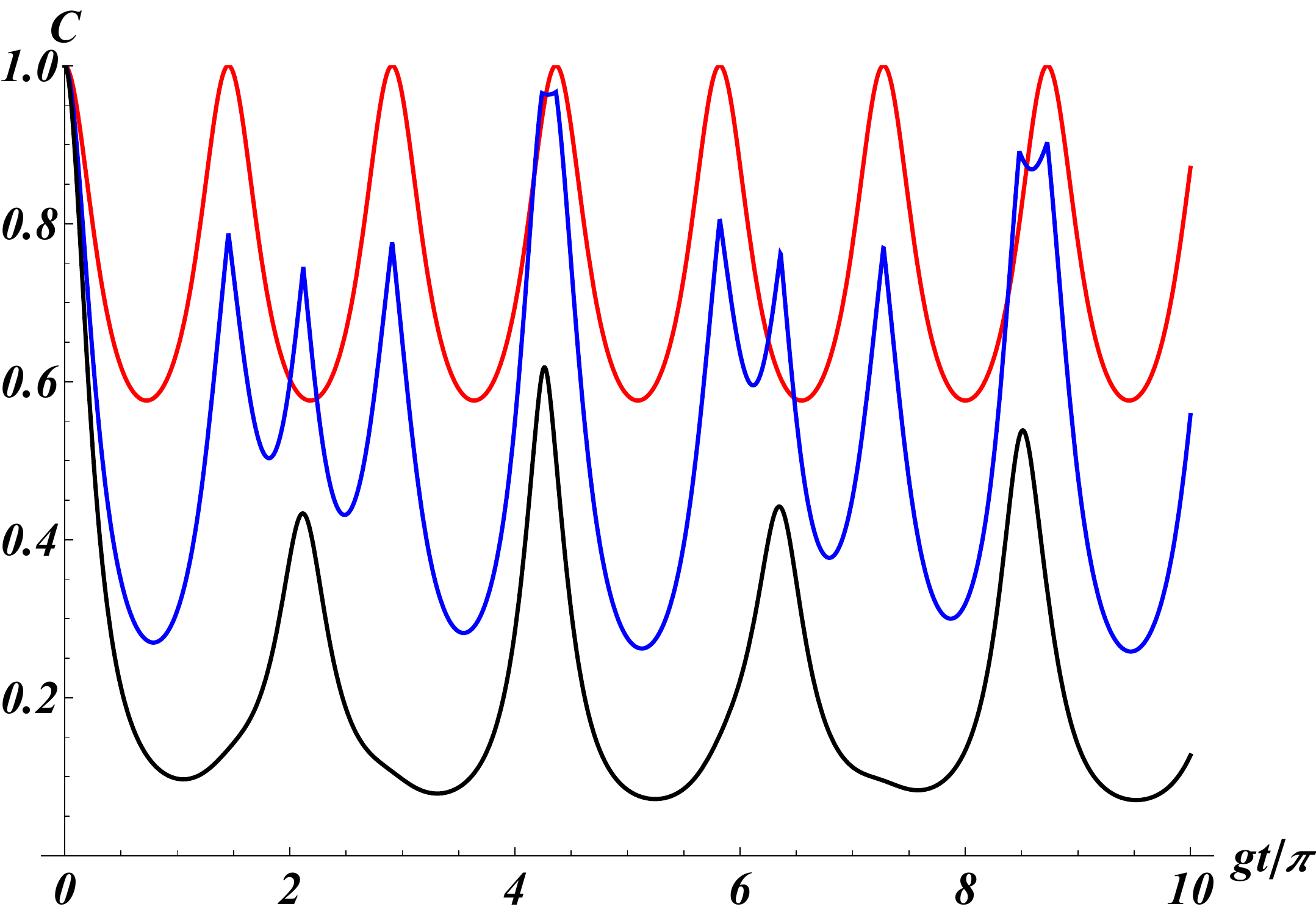}
		\caption{\scriptsize$\kappa=1.7$, $\Omega_3$ is the only frequency in the broken regime.}\label{c}
	\end{subfigure}\hfill
	\begin{subfigure}[b]{0.45\textwidth}
		\includegraphics[scale=0.38]{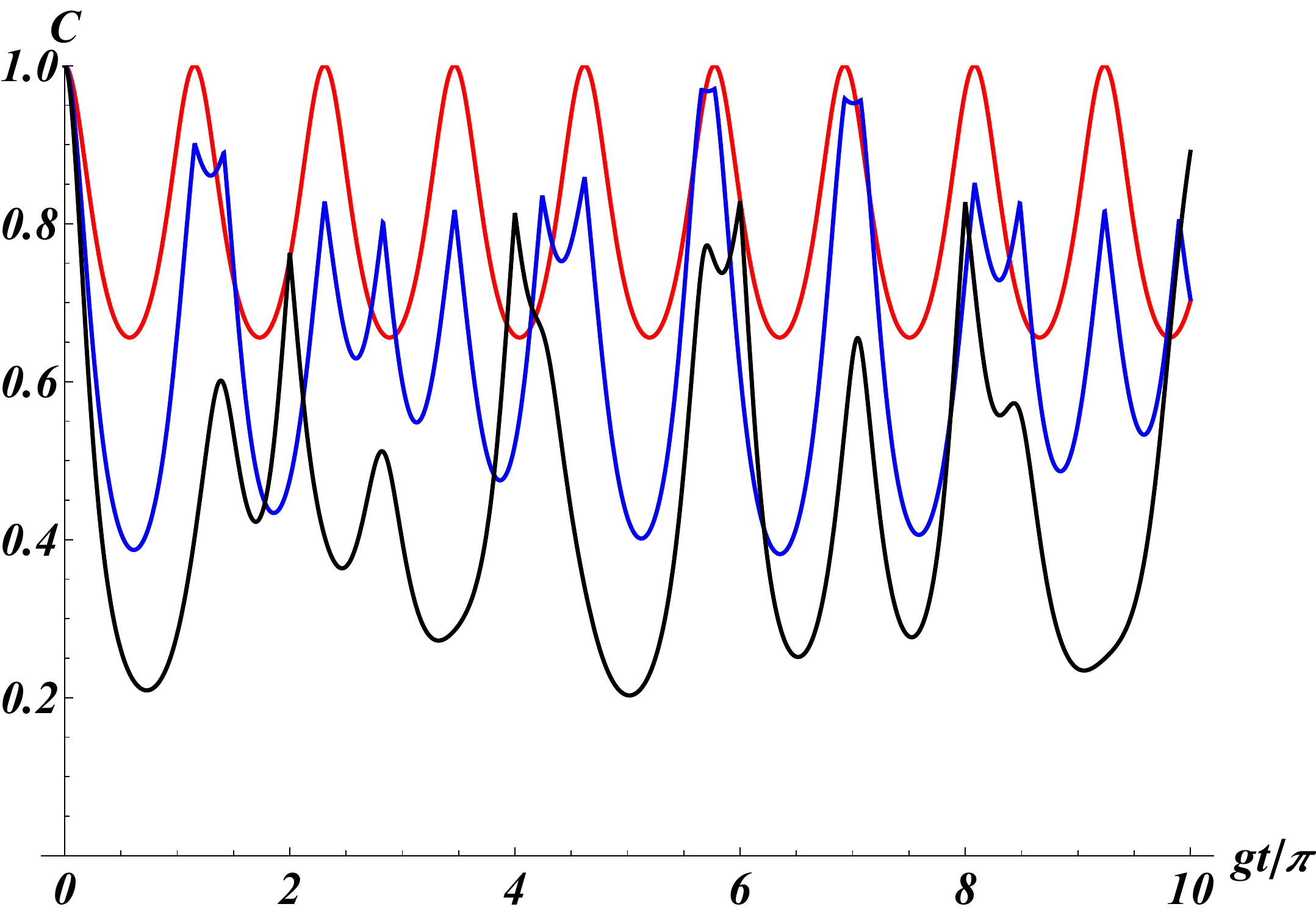}
		\caption{\scriptsize$\kappa=2$, no frequencies are in the broken regime.}\label{d}
	\end{subfigure}

\caption{\small Concurrence $C\left(t\right)$ over time for the state (\ref{state}) with $n=0$ (red), $n=1$ (blue) and $n=2$ (black). Each panel is a plot for a different value of $\kappa=\left(\omega-\nu\right)/g$.}\label{Concurrence}
		
\end{figure}
Figure (\ref{Concurrence}) shows the concurrence for the state (\ref{state}) for $n=0$, $n=1$ and $n=2$ for the values in panels (\ref{a}) $\kappa=0.9$, (\ref{b}) $\kappa=1.4$, (\ref{c}) $\kappa=1.7$ and (\ref{d}) $\kappa=2$. In each case there is a clear change in the evolution of the entanglement including new and exotic behaviour, such as entanglement decaying to a constant value. However, the real discovery is the fact that the entanglement can behave so differently based on the parameters of the Hamiltonian and the initial state of the wavefunction. For example, we can see in panels \ref{a} and \ref{b} that by simply altering $\kappa$ from $0.9$ to $1.4$ we move from decaying to oscillatory behaviour in the states $n=0$ and $n=1$.

This exotic behaviour can be explained by understanding the spontaneous breaking of $\mathcal{PT}$-symmetry in the wavefunction (\ref{state}). The $\mathcal{PT}$-symmetry for $\Omega_m$ is broken when $m\geq\kappa^2$. When this occurs the evolution for this frequency mode changes from trigonometric, to hyperbolic. Meaning we see a shift from oscillatory to decaying evolution. Therefore, when a frequency mode enters the broken regime, it ceases to oscillate and instead decays to a constant. We can calculate this constant, in the broken regime the time-dependent functions approach the following constant

\begin{eqnarray}
|D_n\left(t\rightarrow \infty\right)\delta_n\left(t\rightarrow \infty\right)^{1/2}|&\rightarrow& \frac{1}{\sqrt{2}},\\  |U_n\left(t\rightarrow \infty\right)\delta_n\left(t\rightarrow \infty\right)^{1/2}|&\rightarrow&  \frac{1}{\sqrt{2}}.
\end{eqnarray}
We see this most clearly in panel (\ref{a}) in which all frequencies are in the broken regime. For $n=0$ the concurrence contains only one frequency at $\Omega_1$, and the concurrence decays to a constant 

\begin{equation}
C\left(t\rightarrow\infty\right)\rightarrow\cos\gamma\left(\sqrt{\sin^2\gamma+\frac{1}{4}\cos^2\gamma}-\frac{1}{2}\cos\gamma\right), \quad \text{for} \quad n=0 \quad \text{and} \quad \kappa<1
\end{equation}
For $n>0$ and $\kappa<1$, $C\left(t\rightarrow\infty\right)\rightarrow0$. 

In panel \ref{b} with $1<\kappa<\sqrt{2}$ only one frequency ($\Omega_1$) is left unbroken and so we see oscillations in $n=0$ and $n=1$ of only one frequency mode. However, for $n=1$ the entanglement never returns to its initial value of $1$. Subsequent oscillations only reach $1/\sqrt{2}$ because of the decay of $\Omega_2$. Entanglement in the state with $n=2$ decays to zero but we see a small oscillation (the contirbution from $\Omega_1$) before this happens. 

In panel \ref{c} with $\sqrt{2}<\kappa<\sqrt{3}$ we see the resulting entanglement when $\Omega_3$ is the only frequency in the broken regime. The oscillation of $n=0$ remains unaffected as it only contains $\Omega_1$. We see a change in $n=1$ as expected. With the frequency $\Omega_2$ now unbroken, this now contirbutes as an oscillatory mode. Therefore this state now oscillates with the combination of $\Omega_1$ and $\Omega_2$ with no decay. The state $n=2$ now also oscillates with the combination $\Omega_1$ and $\Omega_2$. However, with the presence of the broken $\Omega_3$ mode the oscillations are damped and entanglement will never reach its initial value.

For the final panel \ref{d} with $\kappa>\sqrt{3}$ all frequency modes are unbroken and so we see oscillation in all three states. For $n=0,1$ the nature of oscillation is unchanged from panel \ref{c}. However, $n=2$ now gains a third oscillatory mode, $\Omega_3$, and so the resulting oscillations are a combination of the three frequencies present in the state $\Omega_1$, $\Omega_2$ and $\Omega_3$. 

Higher values of $n$ will see similar effects to those presented for $n=2$ with transitions in entanglement behaviour at $\kappa=1, \sqrt{n}, \sqrt{n+1}$.

\section{Conclusion}

In this paper we presented, for the first time, time-dependent solutions for the Dyson map and metric operator for the non-Hermitian Jaynes-Cummings Hamiltonian. Until \cite{fring2020spectrally}, time-dependent solutions to other models have always relied on the existence of a closed algebra in order to form an ansatz \cite{fring2016non,fring2018tdm,FRING20172318,HigherDims}, however this was not possible for the JC Hamiltonian. In this paper, we formed an ansatz for the time-dependent Dyson map by taking inspiration from the time-independent case (which we solved starting from perturbation theory).

The solution for the Dyson and metric operator was used to construct the reduced Hermitian density operator and subsequently the concurrence between two isolated non-Hermitian JC Hamiltonians. It was shown that the concurrence exhibits exotic behaviour due to the presence of numerous points of spontaneous $\mathcal{PT}$-symmetry breaking. At these exceptional points, oscillatory evolution transitions to decay and as such we see a drastic shift in the overall behaviour. This is the same effect observed in \cite{fring2019eternal}

The ability to control the evolution of entanglement by varying an overall parameters may have implications for the construction of quantum computers. Preventing the death of entangled states is of utmost importance for the operation of such systems \cite{palma1996quantum,unruh1995maintaining} and so the work presented in this manuscript is highly relevant. The challenge is to build a laboratory system that imitates that of the non-Hermitian system presented here. Non-Hermitian systems have been realised in classical optical experiments e.g. \cite{guo2009observation,ruter2010observation}. Therefore, it is certainly possible that the same could be achieved in quantum computing.

\section*{Acknowledgements}
The author would like to thank Prof. Andreas Fring for his insightful discussions and constructive comments during the initial work and editing process. Furthermore, the author thanks City, University of London for supporting him with a Research Fellowship.

\bibliography{Bibliography}
\bibliographystyle{frithstyle}

\end{document}